\documentclass[twocolumn,showpacs,preprintnumbers]{revtex4}

\usepackage{graphicx}% Include figure files
\usepackage{dcolumn}% Align table columns on decimal point
\usepackage{bm}% bold math

\begin{document}

\title{Dose dependence of ferromagnetism in Co-implanted ZnO}

\author{Numan Akdogan}
 \altaffiliation{Author to whom correspondence should be addressed. E-mail address:
numan.akdogan@ruhr-uni-bochum.de \\
Present address: Department of Physics, Gebze Institute of
Technology, Gebze, 41400 Kocaeli, Turkey}
\author{Hartmut Zabel}
\author{Alexei Nefedov}
 \altaffiliation{Present address: Lehrstuhl f\"{u}r Physikalische Chemie I, Ruhr-Universit\"{a}t Bochum, D-44780 Bochum, Germany}
\author{Kurt Westerholt}
\affiliation{Institut f\"{u}r
Experimentalphysik/Festk\"{o}rperphysik, Ruhr-Universit\"{a}t
Bochum, D-44780 Bochum, Germany}

\author{Hans-Werner Becker}
 \affiliation{Institut f\"{u}r Physik mit Ionenstrahlen, Ruhr-Universit\"{a}t Bochum, D-44780 Bochum, Germany}

\author{\c{S}afak G\"{o}k}
 \affiliation{Lehrstuhl f\"{u}r Angewandte Festk\"{o}rperphysik, Ruhr-Universit\"{a}t Bochum, D-44780 Bochum, Germany}

\author{Rustam Khaibullin}
\author{Lenar Tagirov}
\affiliation{Kazan Physical-Technical Institute of RAS, 420029
Kazan, Russia}
\affiliation{Kazan State University, 420008 Kazan,
Russia}

\date{\today}% It is always \today, today,

\begin{abstract}
We have studied the structural, magnetic and electronic properties
of Co-implanted ZnO(0001) films grown on Al$_2$O$_3$
($11\overline{2}0$) substrates for different implantation doses and
over a wide temperature range. Strong room temperature
ferromagnetism is observed with magnetic parameters depending on the
cobalt implantation dose. A detailed analysis of the structural and
magnetic properties indicates that there are two magnetic phases in
Co-implanted ZnO films. One is a ferromagnetic phase due to the
formation of long range ferromagnetic ordering between implanted
magnetic cobalt ions in the ZnO layer, the second one is a
superparamagnetic phase, which occurs due to the formation of
metallic cobalt clusters in the Al$_2$O$_3$ substrate. Using x-ray
resonant magnetic scattering, the element specific magnetization of
cobalt, oxygen and Zn was investigated. Magnetic dichroism was
observed at the Co \emph{L}$_{2,3}$ edges as well as at the O
\emph{K} edge. In addition, the anomalous Hall effect is also
observed, supporting the intrinsic nature of ferromagnetism in
Co-implanted ZnO films.
\end{abstract}


\maketitle

\section{\label{sec:level1}Introduction}

ZnO is a II-VI semiconductor with a wide band gap of about 3.4 eV.
The stable crystal structure of ZnO is the wurtzite structure
(hexagonal, with $a=3.25~\AA$ and $c=5.12~\AA$)
\cite{WyckoffWiley86}, in which each atom of zinc is surrounded by
four oxygen atoms in tetrahedral coordination. The magnetic
transition metal doped ZnO is interesting from the view point of
forming a transparent ferromagnetic material, and it has the
potential to be a highly multifunctional material with coexisting
ferromagnetic, semiconducting, and magneto-optical properties.
Theoretical predictions of room temperature ferromagnetism in
transition metal (TM)-doped ZnO
\cite{DietlSci00,SatoJJAP00,SatoJJAP01} have initiated a number of
experimental works on these systems as a potential oxide-based
diluted magnetic semiconductor (DMS) material. The first
observation of ferromagnetism in Co-doped ZnO was reported by Ueda
\emph{et al.} \cite{UedaAPL01}. They prepared Zn$_{1-x}$Co$_x$O
thin films on sapphire substrates using PLD technique with $x$
varying between 0.05 and 0.25. Following these initial theoretical
and experimental reports, different growth methods have been used
to deposit Co:ZnO films, including radio-frequency (RF) magnetron
co-sputtering \cite{LimSSC03}, pulsed laser deposition (PLD) using
a KrF laser
\cite{YooJAP01,KimJAP02,KimPB03,PrellierAPL03,VenkatesanPRL04,
YanJAP04,RamachandranAPL04,Ivill08}, combinatorial laser molecular
beam epitaxy (LMBE) \cite{JinAPL01,ZhengJCG05}, sol-gel method
\cite{LeeAPL02}, as well as ion implantation
\cite{AkdoganJPDAP08}. Sapphire has been widely used as substrate
due to the small mismatch (2\%) between (0001) oriented ZnO and
Al$_2$O$_3$ ($11\overline{2}0$) substrates. In addition to cobalt,
other 3\emph{d} transition elements have also been used for
doping, including Mn
\cite{FukumuraAPL99,UedaAPL01,JinAPL01,ChengJAP03,Chakrabarti08},
Ni \cite{UedaAPL01,JinAPL01,VenkatesanPRL04}, V
\cite{JinAPL01,VenkatesanPRL04,SaekiJPC04}, Cr
\cite{UedaAPL01,VenkatesanPRL04}, and also Fe
\cite{UedaAPL01,JinAPL01,ChoAPL02,VenkatesanPRL04}.

Various solubility limits for Co in ZnO were reported by different
groups. Prellier \emph{et al.} \cite{PrellierAPL03} have determined
a solubility limit of about 10 at.\% in PLD-grown films. Park
\emph{et al.} \cite{ParkAPL04} reported that cobalt nanoclusters
start to form for $x\geq$12 at.\% in samples grown by sol-gel and RF
sputtering techniques. Lee \emph{et al.} \cite{LeeAPL02} observed
some undefined Bragg peaks for a cobalt concentrations higher than
25 at.\%. Kim \emph{et al.} \cite{KimJAP02} showed that the
solubility limit is less than 40 at.\% in PLD-grown films. Ueda
\emph{et al.} \cite{UedaAPL01} claimed that the solubility limit is
lower than 50 at.\% and they clearly observed a phase separation
into ZnO- and CoO-rich phases in the film prepared using
Zn$_{0.5}$Co$_{0.5}$O targets. These controversial results from
different research groups are likely due to different growth
techniques used and/or due to different growth conditions such as
oxygen pressure and deposition temperature. Recently, we have
reported that using ion implantation cobalt concentrations of up to
50 at.\% in ZnO are possible without cobalt cluster formation
\cite{AkdoganJPDAP08}. This high concentration is attributed to the
properties of ion implantation, which allows doping of transition
metals beyond their equilibrium solubility limits
\cite{HebardJPD04}.

Regarding the magnetic properties of Co-doped ZnO films, while
several groups including ourself have observed room temperature
ferromagnetism for 50 at.\% \cite{AkdoganJPDAP08}, 25-30 at.\%
\cite{LeeAPL02,Ivill08} and lower
\cite{PrellierAPL03,VenkatesanPRL04,RamachandranAPL04,YinJAP04} Co
concentrations, others reported the absence of ferromagnetism at
room temperature \cite{JinAPL01,KimJAP02,ParkAPL04}.

In this paper we report detailed studies using various experimental
techniques for the investigation of the structural, magnetic and
electronic properties of Co-implanted ZnO films grown on sapphire
substrates  and for different cobalt concentrations. Rutherford
backscattering spectroscopy (RBS) and X-ray diffraction (XRD) were
used to determine the depth profile of implanted cobalt ions and to
detect the formation of possible secondary phases such as metallic
cobalt clusters. The magnetic properties of the films were
characterized by the magneto-optical Kerr effect (MOKE), a
superconducting quantum interference device (SQUID) magnetometer, as
well as x-ray resonant magnetic scattering (XRMS) techniques. In
order to determine the type and concentration of carriers in
Co-implanted ZnO films, Hall effect measurements were also
performed.

\section{\label{sec:level1}Sample Preparation}

\begin{figure} \centering
\includegraphics[clip=true,keepaspectratio=true,width=\linewidth]
{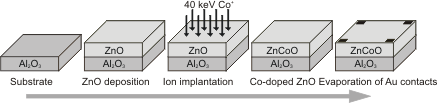} \caption{\label{Sample-ZnO} Sample preparation
stages for Co-implanted ZnO/Al$_2$O$_3$ films.}
\end{figure}

About 35 nm thick ZnO(0001) films were grown on $10\times10$
mm$^{2}$ epi-polished single-crystalline Al$_2$O$_3$
($11\overline{2}0$) substrates by RF (13.56 MHz) sputtering of a ZnO
target \cite{AyASS03}. The sputtering was carried out in an
atmosphere of $5\times10^{-3}$ mbar pure Ar ($99.999 \%$) with a
substrate temperature of $500^\circ C$. In order to increase the
quality of ZnO films, the samples were annealed in an oxygen
atmosphere with a partial pressure of up to 2000 mbar and a
temperature of $800^\circ C$. After annealing, the ZnO samples were
implanted with 40 keV Co$^+$ ions with an ion current density of $8
\mu A\cdot cm^{-2}$ using the ILU-3 ion accelerator (Kazan
Physical-Technical Institute of Russian Academy of Science). The
sample holder was cooled by flowing water during the implantation to
prevent the samples from overheating. The implantation dose varied
in the range of $0.25-2.00\times10^{17} ions\cdot cm^{-2}$. After
implantation, the samples were cut into square pieces and gold
contacts were evaporated on the corners of the samples for Hall
effect studies (Fig.~\ref{Sample-ZnO}). A list of the Co-implanted
ZnO films used for the present study is given in
Table~\ref{ZnO-samples}.

\begin{table}[!h]
\caption{List of the ZnO films implanted with 40 keV Co$^+$ for
varying Co ion dose.} \centering
\begin{tabular}{cc}
\hline
Sample & Dose ($\times10^{17} ion\cdot cm^{-2}$) \\
\hline
1 & 0.25 \\
2 & 0.50 \\
3 & 0.75 \\
4 & 1.00 \\
5 & 1.25 \\
6 & 1.50 \\
7 & 2.00 \\
\hline
\end{tabular}
\label{ZnO-samples}
\end{table}

\section{\label{sec:level1}Experimental Results}

\subsection{\label{sec:level2}Structural Properties}

\begin{figure}[!h]
\includegraphics[width=0.5\textwidth]{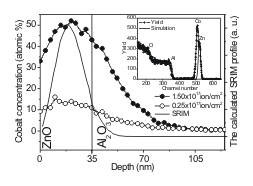}
\caption{\label{RBS-SRIM-ZnO} Depth dependence of the cobalt
concentration in ZnO/Al$_2$O$_3$ implanted with Co ions with a dose
of $0.25\times10^{17} ions\cdot cm^{-2}$ (open symbols) and
$1.50\times10^{17} ions\cdot cm^{-2}$ (black symbols), respectively.
Solid line presents the calculated SRIM profile. The inset shows the
experimentally observed (symbols) and simulated (solid line) RBS
spectra for sample 6.}
\end{figure}

The depth dependence of the cobalt concentration in Co-implanted
ZnO/Al$_2$O$_3$ films was investigated using the RBS technique at
the Dynamic Tandem Laboratory (DTL) at the Ruhr-Universit\"{a}t
Bochum. The RBS data show both a maximum of cobalt concentration
located close to the ZnO/Al$_2$O$_3$ interface and an extended
inward tail due to cobalt diffusion into the volume of the
Al$_2$O$_3$ substrate (Fig.~\ref{RBS-SRIM-ZnO}). We also noticed
that after ion implantation the thickness of the ZnO layer has
shrunk (e.g., from originally 35 nm to 28 nm for sample 6) due to
sputtering effects. According to the SRIM algorithm
\cite{ZieglerPP85}, the average implanted depth of 40 keV Co ions in
ZnO/Al$_2$O$_3$ is about 20.4 nm with a straggling of 9.6 nm in the
Gaussian-like depth distribution (solid line in
Fig.~\ref{RBS-SRIM-ZnO}). However, because of the surface
sputtering, ion mixing and heating of the implanted region by the
ion beam, a redistribution of the implanted cobalt ions compared to
the calculated profile has to be taken into account.

\begin{figure}[!h]
\includegraphics[width=0.5\textwidth]{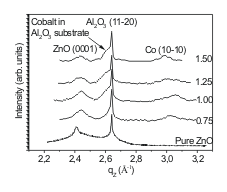}
\caption{\label{Xrd-ZnO} High angle Bragg scans
of the ZnO(0001) films on Al$_2$O$_3$
($11\overline{2}0$) before and after cobalt ion
implantation.}
\end{figure}

High-angle XRD experiments provide information on
the structural coherence of the films and in our
case also of possible additional phases in the
sample after ion implantation. Fig.~\ref{Xrd-ZnO}
shows high angle Bragg scans of the ZnO films
before and after cobalt implantation. The data
were taken using synchrotron radiation at the
"Hamburg Synchrotron Radiation Laboratory"
(HASYLAB) (for pure ZnO film) and at the
"Dortmund Electron Accelerator" (DELTA) (for
cobalt implanted ZnO films) with an energy of
E=11000 eV. Before implantation the x-ray
diffraction pattern consists of a very strong
Al$_2$O$_3$ ($11\overline{2}0$) peak and a
ZnO(0001) reflection to the left side. The ZnO
peak is surround by thin film Laue oscillations,
which are indicative for the high quality of the
ZnO film. After implantation, the XRD diffraction
pattern shows a ($10\overline{1}0$) reflection of
the Co hcp structure on the right side of the
sapphire substrate peak. The ion bombardment also
causes an intensity reduction of the ZnO(0001)
peak proportional to the implantation dose,
indicating an increasing amount of lattice
defects. Furthermore, after implantation we
observe a shift of the ZnO (0001) peak to higher
angles. We attribute the ZnO lattice contraction
to the substitution of Zn ions by Co cobalt ions,
which has a smaller ion radius. In addition,
after implantation a tail (shown by an arrow in
Fig.~\ref{Xrd-ZnO}) appears on the low angle side
of the main Al$_2$O$_3$ ($11\overline{2}0$) peak
which is not observed before implantation. This
tail likely reflects the lattice expansion of the
sapphire substrate upon Co implantation.

In addition to  the XRD experiments, we have also
performed high resolution cross sectional
transmission electron microscopy (TEM)
measurements for sample 6 \cite{AkdoganJPDAP08}.
The TEM results reveal the presence of metallic
cobalt clusters in the Al$_2$O$_3$ sapphire
substrate, but not in the ZnO film. Co clusters
with a diameter of about 5-6 nm form a Co rich
layer in the substrate close to the
ZnO/Al$_2$O$_3$ interface \cite{AkdoganJPDAP08}.

\subsection{\label{sec:level2} Magnetic Properties}

\subsubsection{\label{sec:level3} Room temperature magnetization measurements}

\begin{figure}[!h]
\centering
\includegraphics[width=0.4\textwidth]{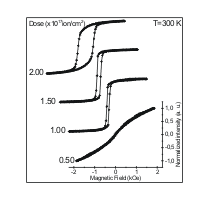}
\caption{\label{Moke-Hys-ZnO} Room temperature MOKE hysteresis
curves of Co-implanted ZnO films measured for varying implantation
dose.}
\end{figure}

For the investigation of the magnetic properties of the Co implanted
samples we used a high-resolution MOKE setup in the longitudinal
configuration with s-polarized light
\cite{ZeidlerPRB96,SchmitteJAP02,WestphalenRSI07}.
Fig.~\ref{Moke-Hys-ZnO} shows the hysteresis loops of Co-implanted
ZnO films which were recorded at room temperature. The MOKE data in
Fig.~\ref{Moke-Hys-ZnO} clearly indicate that after cobalt
implantation, non-magnetic ZnO becomes ferromagnetic at room
temperature with a large remanent magnetization. With increasing
cobalt concentration the implanted ZnO films exhibit sequentially
paramagnetic, weak ferromagnetic and, finally, ferromagnetic
response with a square-like hysteresis at room temperature for the
dose of $1.50\times10^{17} ions\cdot cm^{-2}$. For the highest dose
($2.00\times10^{17} ions\cdot cm^{-2}$) the square-like shape of the
hysteresis loop disappears and the coercive field increases
drastically. From this we infer that for the highest dose level the
cobalt atoms start to form clusters in the ZnO film. Moreover,
although no in-plane magnetic anisotropy was observed by MOKE in
Co-implanted ZnO films, we observed a clear six-fold in-plane
magnetic anisotropy by ferromagnetic resonance (FMR) technique
\cite{AkdoganFMR08}. The corresponding FMR data show that the easy
and hard axes have a periodicity of 60 degree in the film plane, in
agreement with the hexagonal structure of the ZnO film.

In order to study in  detail the observed
ferromagnetic behavior, the magnetic properties
of Co-implanted ZnO films were investigated using
the XRMS technique. XRMS has proven to be a
highly effective method for the analysis of the
magnetic properties of buried layers and
interfaces, including their depth dependence
\cite{TonnerrePRL95,LaanCOSSMS06}. Moreover, if
the photon energy is fixed close to the energy of
the corresponding x-ray absorption edges, element
specific hysteresis loops can be measured
\cite{KortrightNIMB03}. Since there are three
elements in the Co-doped ZnO film, the analysis
can be carried out separately for Co, O and Zn.

The XRMS experiments  were performed using the ALICE diffractometer
\cite{GrabisRSI03} at the undulator beamline UE56/1-PGM at BESSY II
(Berlin, Germany). The diffractometer comprises a two-circle
goniometer and works in horizontal scattering geometry. A magnetic
field can be applied in the scattering plane and along the sample
surface either parallel or antiparallel to the photon helicity,
which corresponds to the longitudinal magneto-optical Kerr effect
(L-MOKE) geometry. The maximum field of $\pm2700$ Oe was high enough
to fully saturate the magnetization of the sample. The magnetic
contribution to the scattered intensity was always measured by
reversing the magnetic field at fixed photon helicity. As a
compromise between high scattering intensity and high magnetic
sensitivity for the investigation of the magnetic properties at the
Co \emph{L} edges, the scattering angle was fixed at the position of
$2\theta=8.2^\circ$ (the angle of incidence is $\theta=4.1^\circ$)
\cite{AkdoganJPDAP08}.

\begin{figure}[!h]
\centering
\includegraphics[width=0.4\textwidth]{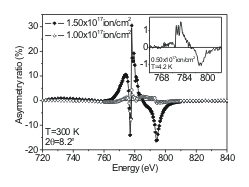}
\caption{\label{Xrms-Asym-Co-ZnO} The asymmetry ratios taken at the
Co \emph{L} edges for sample 6 ($1.50\times10^{17} ions\cdot
cm^{-2}$) and sample 4 ($1.00\times10^{17} ions\cdot cm^{-2}$) shown
by black and open symbols, respectively. Inset presents the
asymmetry ratio of sample 2 ($0.50\times10^{17} ions\cdot cm^{-2}$)
measured at 4.2 K.}
\end{figure}

The  magnetic contribution to the resonant scattering can best be
visualized by plotting the asymmetry ratio,
$A_r=(I^+-I^-)/(I^++I^-)$.  In Fig.~\ref{Xrms-Asym-Co-ZnO} we show
the asymmetry ratio taken at the Co \emph{L} edges for samples doped
with different doses. The asymmetry ratio shows a strong
ferromagnetic signal for sample 6 (up to 30 \%), and it decreases
with decreasing cobalt implantation dose. For sample 2, we observe
only a very small magnetic signal at 4.2 K. In addition to XRMS, we
have also employed x-ray absorption spectroscopy (XAS) experiments
for sample 6. The XAS spectrum clearly exhibits a multiplet
structure of the Co $\emph{L}_3$ peak, which is typical for oxidized
cobalt showing the presence of Co$^{2+}$ state in the ZnO film
\cite{AkdoganJPDAP08}.

\begin{figure}[!h]
\centering
\includegraphics[width=0.5\textwidth]{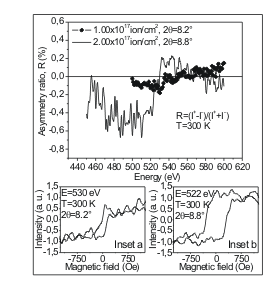}
\caption{\label{Xrms-Asym-O-ZnO}  The magnetic
signal at the O \emph{K} edge for samples 4
(black symbols) and 7 (solid line). Insets a and
b show the hysteresis curves taken at the O
\emph{K} edge for samples 4 and 7, respectively.}
\end{figure}

The magnetic signal at  the Zn $\emph{L}_3$-
(E=1021.8 eV) and the O \emph{K}- (526.8 eV)
edges were also investigated. Within the
sensitivity limit no magnetic signal could be
resolved for Zn. However, a clear magnetic signal
was observed at the O \emph{K} edge for sample 6
\cite{AkdoganJPDAP08}. In addition to sample 6, a
very small magnetic signal at the O \emph{K} edge
was also observed for the samples 4 and 7
presented in Fig.~\ref{Xrms-Asym-O-ZnO}.

\subsubsection{\label{sec:level3} Temperature dependent magnetization measurements}

\begin{figure}[!h]
\centering
\includegraphics[width=0.4\textwidth]{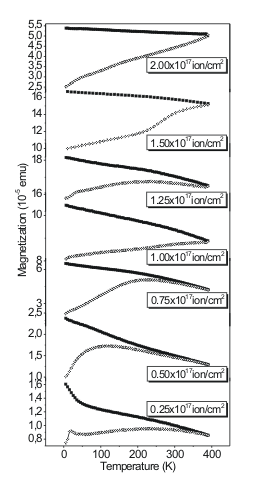}
\caption{\label{Squid-M-T-ZnO} Temperature dependent magnetization
curves of Co-implanted ZnO films recorded by SQUID magnetometry
for varying implantation dose. FC and ZFC curves refer to field
cooled and zero-field cooled protocols and are presented by closed
and open symbols, respectively. In both cases the data were taken
in a field of 100 Oe during the heating up cycle.}
\end{figure}

\begin{figure}[!h]
\centering
\includegraphics[width=0.4\textwidth]{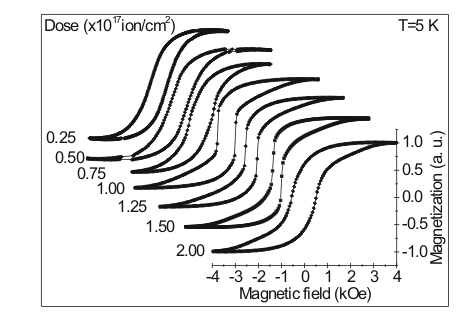}
\caption{\label{Squid-Hys-5K-ZnO} SQUID $M-H$ loops of Co-implanted
ZnO films measured for different implantation doses at 5 K.}
\end{figure}

\begin{figure}[!h]
\centering
\includegraphics[width=0.4\textwidth]{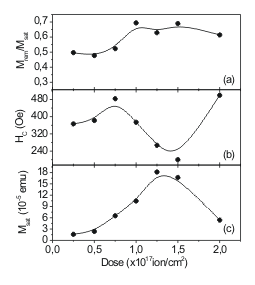}
\caption{\label{Squid-Dose-5K-ZnO} The dose dependence of the
normalized remanent magnetization (a), the coercive field (b) and
the saturation magnetization (c), respectively. The data taken at 5
K using SQUID magnetometry.}
\end{figure}

In order to check the temperature  dependence of the magnetization
for ZnO films doped with different doses, we carried out field
cooled (FC) and zero field cooled (ZFC) $M-T$ measurements using a
SQUID magnetometer. For ZFC measurements, the samples are first
cooled in zero field to 5 K and the magnetization is recorded
during warming up to 390 K with an applied field of 100 Oe,
parallel to the  film surface. For FC measurements, the applied
field of 100 Oe is kept constant during cooling to 5 K and the
magnetization is recorded during warming at the same field value.
Due to the clustering of cobalt in the Al$_2$O$_3$ substrate
(\cite{AkdoganJPDAP08}), the FC (closed symbols) and ZFC (open
symbols) curves presented in Fig.~\ref{Squid-M-T-ZnO} always show
evidence for the presence of a superparamagnetic phase. There is a
small peak at about 20 K in ZFC curve of sample 1
($0.25\times10^{17} ions\cdot cm^{-2}$) and this peak shifts to
higher temperatures with increasing cobalt concentration. The
trend in the $M-T$ curve of sample 1 ($0.25\times10^{17} ions\cdot
cm^{-2}$) can be attributed to the coexistence of a ferromagnetic
phase originating from substituted Co$^{2+}$ ions in ZnO and the
superparamagnetic phase due to cluster formation in Al$_2$O$_3$.
Hysteresis curves measured at 5 K (Fig.~\ref{Squid-Hys-5K-ZnO})
indicate that the superparamagnetic phase in this sample is more
dominant than the ferromagnetic phase. The $M-T$ measurements for
the samples implanted with higher doses exhibits
superparamagnetism with a blocking temperature of about 100 K and
250 K for sample 2 ($0.50\times10^{17} ions\cdot cm^{-2}$) and
sample 3 ($0.75\times10^{17} ions\cdot cm^{-2}$), respectively.
The hysteresis curves of these films (Fig.~\ref{Squid-Hys-5K-ZnO})
also show that the superparamagnetic phase is still dominating
over the ferromagnetic phase. But the steep part of the hysteresis
curve of sample 3 and the increased coercivity ($0.75\times10^{17}
ions\cdot cm^{-2}$) are indicative for the onset of a clear
ferromagnetism phase at this dose. The temperature dependent
magnetization curves of sample 4 ($1.00\times10^{17} ions\cdot
cm^{-2}$), sample 5 ($1.25\times10^{17} ions\cdot cm^{-2}$) and
sample 6 ($1.50\times10^{17} ions\cdot cm^{-2}$) show that these
samples have a blocking temperature of about 390 K or even higher.
The  magnetic hysteresis of these samples measured by SQUID
(Fig.~\ref{Squid-Hys-5K-ZnO}) clearly show a ferromagnetic phase
superimposed by a superparamagnetic component.  The ferromagnetic
component is present even above room temperature as seen in the
MOKE experiments in Fig.~\ref{Moke-Hys-ZnO}. Since MOKE probes
only films near their surface, the superparamagnetic component in
these samples, which is deeper in the substrate, is not seen by
MOKE experiments.

In the SQUID hysteresis curves there is another remarkable effect
of the ferromagnetic phase as a function of dose. The coercivity
$H_C$ decreases systematically with increasing Co dose up until a
dose of $1.50\times10^{17} ions\cdot cm^{-2}$, as seen in
Figs.~\ref{Squid-Hys-5K-ZnO} and ~\ref{Squid-Dose-5K-ZnO} (b).
This behavior may be explained as follows: with increasing Co dose
the magnetization becomes more homogeneous and, since magnetic
inhomogeneities are the main source of pinning for the domain
walls, $H_C$ decreases with increasing Co dose. Between
$1.25\times10^{17} ions\cdot cm^{-2}$ and $1.50\times10^{17}
ions\cdot cm^{-2}$ the potential barrier for reversal of the
ferromagnetic component becomes smaller. Up to this level all
inhomogeneities are filled. Any higher dose is counterproductive,
it decreases the saturation magnetization and enhances the
coercivity (see Figs.~\ref{Squid-Dose-5K-ZnO} (b) and (c)),
indicating that Co goes into antisites with eventually
antiferromagnetic (AF) coupling, loss of magnetization, and
increase of the coercivity. CoO clusters are formed in the ZnO
matrix with AF spin structure and AF coupling to the remaining
ferromagnetic Zn(Co)O film. The $M-T$ data
(Fig.~\ref{Squid-M-T-ZnO}) and the room temperature
(Fig.~\ref{Moke-Hys-ZnO}) and low temperature
(Fig.~\ref{Squid-Hys-5K-ZnO}) hysteresis measurements of sample 7
($2.00\times10^{17} ions\cdot cm^{-2}$) clearly indicate that the
cobalt atoms start to cluster also within the ZnO layer at the
highest dose.

\subsection{\label{sec:level2}Hall effect measurements}

In ferromagnetic materials the Hall voltage
consists of the ordinary term and an additional
term that contributes to the Hall voltage due to
their spontaneous magnetization. This additional
contribution, called anomalous Hall effect, is
proportional to the sample magnetization
\cite{HurdPP72}. Hence, the Hall voltage can be
written as \cite{HurdPP72},

\begin{center}
\begin{equation}\label{Hallvoltage}
V_H = \Big(\frac{R_0I}{t}\Big)Hcos\alpha +
\Big(\frac{R_A\mu_0I}{t}\Big)Mcos\theta,
\end{equation}
\end{center}

where $t$ is the film  thickness and $I$ is the
current. $R_0$ and $R_A$ are the ordinary and
anomalous Hall effect coefficients, respectively.
$\mu_0$ is the permeability of free space.
$\alpha$ is the angle between the applied
magnetic field ($H$) and sample normal. $\theta$
is the angle between the sample magnetization
($M$) and the sample normal. The first term in
Eq. \ref{Hallvoltage} is the ordinary Hall effect
and arises from the Lorentz force acting on
conduction electrons. This establishes an
electric field perpendicular to the applied
magnetic field and to the current. The anomalous
Hall effect term is conventionally attributed to
spin dependent scattering involving a spin-orbit
interaction between the conduction electrons and
the magnetic moments of the material. At low
applied magnetic fields, the Hall voltage ($V_H$)
is dominated by the magnetic field dependence of
the sample magnetization $M$. When the applied
magnetic field is high enough to saturate the
sample magnetization, the magnetic field
dependence of the Hall voltage becomes linear due
to the ordinary Hall effect.

\begin{figure}[!h]
\centering
\includegraphics[width=0.4\textwidth]{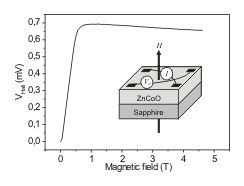}
\caption{\label{Ahe-ZnO} AHE data of sample 6 ($1.50\times10^{17}
ions\cdot cm^{-2}$) taken at 4.2 K. Inset shows the geometry of the
AHE measurements. $H$ is the external magnetic field applied
perpendicular to the film surface.}
\end{figure}

In order to check whether  this behavior is
present in Co-implanted ZnO films and to
determine the character of the majority carriers,
we have carried out Hall effect experiments. The
Hall effect measurements were performed at 4.2 K
using a van der Pauw configuration presented in
Fig.~\ref{Ahe-ZnO} as an inset.

The Hall effect data of  sample 6 ($1.50\times10^{17} ions\cdot
cm^{-2}$) are shown in Fig.~\ref{Ahe-ZnO}. A sharp rise in the
Hall voltage at low field, i.e., AHE, is followed by a slow
decrease corresponding to the ordinary Hall effect. It is
important to note that the negative slope at high fields indicates
\emph{n}-type carriers in Co-implanted ZnO film with a 3D carrier
concentration of $n_{3D}=1.931\times10^{19}\cdot cm^{-3}$. The
Hall mobility measured at 4.2 K is about $90 cm^2 \cdot
V^{-1}s^{-1}$ for sample 6. We have also observed similar behavior
for the samples 3, 4, 5 and 7. However, for the lowest two doses
(samples 1 and 2), the measurements cannot be done because of a
too small signal-to-noise ratio of the Hall voltage.

\section{\label{sec:level1} Discussion}

For the dose dependence of magnetic phases in ZnO films at room
temperature we propose the following scenario : At low doses
($0.25-0.50\times10^{17} ions\cdot cm^{-2}$) the number of
substituted cobalt ions in the ZnO layer is very small, which
results in a paramagnetic signal at room temperature. Increasing
of cobalt implantation dose leads to an increasing number of
substituted cobalt ions and after certain cobalt concentration
they start to interact ferromagnetically. For this reason at the
cobalt dose of $0.75\times10^{17} ions\cdot cm^{-2}$ a weak
ferromagnetic behavior is observed with a $T_c$ below room
temperature. At higher cobalt concentrations
($1.00-1.50\times10^{17} ions\cdot cm^{-2}$) the substituted
cobalt ions in ZnO interact strongly and stabilize room
temperature ferromagnetism. At the highest dose of
$2.00\times10^{17} ions\cdot cm^{-2}$, in addition to the
substituted cobalt ions, metallic cobalt clusters are also present
in the ZnO layer.

\begin{figure}[!h]
\centering
\includegraphics[width=0.25\textwidth]{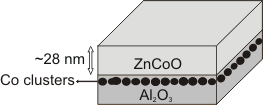}
\caption{\label{Sample-clusters-ZnO} The cluster formation in
Al$_2$O$_3$ substrate after cobalt ion implantation.}
\end{figure}

As discussed in detail in  Ref. \cite{AkdoganJPDAP08}, the
difference in the shape of the hysteresis loops obtained by MOKE
and SQUID is attributed to the surface sensitivity of the MOKE
technique with a maximum penetration depth of about 20-30 nm. The
ZnO films have a thickness of 35 nm before implantation. Because
of surface sputtering, the ZnO thickness decreases (e.g.,
decreased to about 28 nm for sample 6) after implantation. Thus,
MOKE provides information only from the ZnO layer, not from the
sapphire substrate, i.e. MOKE is only sensitive to the
ferromagnetic contribution from the ZnO layer. In this layer a
small fraction of nonmagnetic Zn atoms are substituted by magnetic
Co ions, giving raise to the MOKE hysteresis. However, SQUID
measurements collect magnetic contributions from both the
Co-implanted ZnO film and from the cobalt clusters in Al$_2$O$_3$
(Fig.~\ref{Sample-clusters-ZnO}). Therefore, the difference
between the MOKE and SQUID data appear as a result of the
depth-dependent Co content in the implanted layer.

Another important result of this study is the observation of oxygen
spin polarization in Co-implanted ZnO films. This shows that the
oxygen atoms are polarized due to the spontaneous ferromagnetic
order in ZnO films. The main question that arises here is the
mechanism which leads to the observed long range ferromagnetic
ordering in Co-doped ZnO. Recently, Patterson \cite{PattersonPRB06}
calculated the electronic band structure of Co substituted for Zn in
ZnO, for Zn and O vacancies, and for interstitial Zn in ZnO using
the B3LYP hybrid density functional theory. He reported that the
singly-positively charged O vacancy is the only defect in Co-doped
ZnO which can mediate ferromagnetic exchange coupling between Co
ions at intermediate range (just beyond near neighbor distances). In
the ground state configuration the majority Co spins are parallel,
whereas the minority spins are parallel to each other and to the
oxygen vacancy spin, so that there are exchange couplings between
these three spins leading to an overall ferromagnetic ground state
of the Co ions. No substantial exchange coupling was found for the
positively charged interstitial Zn defect which has also spin 1/2.
The exchange coupling mechanism proposed by Patterson is essentially
the same as the impurity band model of Coey \emph{et al.}
\cite{CoeyNature05}, where the polarons bound to the oxygen
vacancies mediate ferromagnetic coupling between Co ions. In order
to have the magnetic moments of the Co ions aligned
ferromagnetically, one mediating electron is required with an
oppositely directed spin. This is in line with a recent comparison
of band structure calculations by Walsh \textit{et al.} showing that
the electronic structure of Co-doped ZnO is consistent with carrier
mediated ferromagnetism \cite{Walsh08}. The oxygen spin polarization
has not explicitly been considered in the aforementioned band
structure calculations and may be due to ferromagnetic splitting of
nearest neighbor oxygen \emph{p}-levels. This has already been
speculated by Methfessel and Mattis in their seminal review article
on magnetic semiconductors \cite{Methfessel68}.

The reason for the observation of  AHE and
\emph{n}-type carriers in Co-implanted ZnO films
can be explained by electron doping via Zn
interstitials. Normally, isovalent TM$^{2+}$
doping of ZnO does not introduce charge carriers
itself, they need to be produced by additional
doping \cite{NielsenPSS06}. However, using ion
implantation not only cobalt ions are introduced
in ZnO, but simultaneously many other defects are
also be produced in the implanted region, such as
Zn interstitials which are reported to form
shallow donors in ZnO
\cite{LookSSC98,LookPRL99,LeeAPL02}. This can be
thought of as an added advantage of ion
implantation that it not only introduces
transition metal ions to induce ferromagnetism
but also introduces  the required charge carriers
into the ZnO.

\section{\label{sec:level1}Summary}

In conclusion, the structural, magnetic  and
electronic properties of Co-implanted ZnO films,
deposited by RF-sputtering methods on a
($11\overline{2}0$) oriented sapphire substrate,
have been investigated. The structural data
indicate a Co cluster formation in the sapphire
substrate close to the ZnO/Al$_2$O$_3$ interface
but well separated from the ZnO film. No
indication of clustering in the ZnO layer has
been found. The previously reported XAS data with
a multiplet fine structure around the Co
\emph{L}$_3$ edge clearly shows that the
implanted cobalt ions are in the Co$^{2+}$
oxidation state, most likely substituting part of
the Zn$^{2+}$ ions in the host matrix. The
combination of room temperature and low
temperature magnetization measurements indicates
that there are two magnetic phases in the
Co-implanted ZnO/Al$_2$O$_3$ films. One is the
ferromagnetic phase due to the Co substitution on
Zn sites in the ZnO film, the second magnetic
phase originates from Co clusters in the sapphire
substrate. Furthermore, a clear ferromagnetic
signal at the O \emph{K} edge is observed which
shows that the oxygen spin polarization is an
important indicator for the observed long range
ferromagnetic ordering in the ZnO layer. In
conclusion, implantation of cobalt ions into the
nonmagnetic ZnO film causes intrinsic
ferromagnetism at room temperature and
simultaneously creates \emph{n}-type charge
carriers without additional doping.

\begin{acknowledgments}
We would like to acknowledge S. Erdt-B\"{o}hm and P. Stauche for
sample preparation and technical support.  We also would like to
thank also Dr. C. Sternemann and Dr. M. Paulus for their assistance
with the beamline operation at DELTA, and G. Nowak for his help to
perform XRD experiments at HASYLAB. This work was partially
supported by BMBF through Contracts Nos. 05KS4PCA (ALICE Chamber)
and 05ES3XBA/5 (Travel to BESSY), by DFG through SFB 491, and by
RFBR through the grant Nos 07-02-00559-a and 04-02-97505-r. N.
Akdogan acknowledges a fellowship through the International Max
Planck Research School-SurMat.
\end{acknowledgments}

%\newpage %Just because of unusual number of tables stacked at end
\bibliography{ZnO2}% Produces the bibliography via BibTeX.

\end{document}